\documentclass[
superscriptaddress,
reprint,
amsmath,amssymb,
aps,
prl,
floatfix,
longbibliography,
twocolumn
]{revtex4-2}

\usepackage{mathrsfs}
\usepackage{graphicx}
\usepackage{dcolumn}
\usepackage{bm}
\usepackage{chemformula}
\usepackage{braket}
\usepackage{multirow}
\usepackage{booktabs}
\usepackage{tabularx}  
\usepackage{adjustbox}  
\usepackage{hyperref}
\usepackage{natbib}
\hypersetup{
    colorlinks=true,
    linkcolor=blue,
    citecolor=blue,
    urlcolor=blue,
}

\usepackage{xr}
\begin{document}

\title{Switchable Altermagnetism via Spin-Induced Improper Polarization}

\author{Zhihao Dai}
\affiliation{Key Laboratory of Computational Physical Sciences (Ministry of Education), Institute of Computational Physical Sciences, State Key Laboratory of Surface Physics and Department of Physics, Fudan University, Shanghai 200433, China}

\author{Michele Reticcioli}
\affiliation{CNR-SPIN, c/o Dip.to di Scienze Fisiche e Chimiche - Via Vetoio - 67100 - Coppito (AQ), Italy}
\affiliation{TU Wien, Institute of Applied Physics, Wiedner Hauptstr. 8-10/134, 1040 Vienna, Austria}

\author{Wei Ren}
\affiliation{Physics Department, Materials Genome Institute,
Institute for Quantum Science and Technology, Shanghai University, Shanghai 200444, China}

\author{Hongjun Xiang}
\email{hxiang@fudan.edu.cn}
\affiliation{Key Laboratory of Computational Physical Sciences (Ministry of Education), Institute of Computational Physical Sciences, State Key Laboratory of Surface Physics and Department of Physics, Fudan University, Shanghai 200433, China}

\author{Alessandro Stroppa}
\email{alessandro.stroppa@spin.cnr.it}
\affiliation{CNR-SPIN, c/o Dip.to di Scienze Fisiche e Chimiche - Via Vetoio - 67100 - Coppito (AQ), Italy}

\begin{abstract}
Enabling reversible spin-splitting switching in stray-field-free altermagnets is promising for spintronic applications, but currently limited to a narrow class of polar materials.
We propose a broader approach based on spin-induced improper polarization in nonpolar dual-sublattice magnets.
We demonstrate this mechanism in \ch{DyFeO3}, where the product of nonpolar \ch{Fe} and \ch{Dy} spin modes transforms as an induced polar mode.
Density functional theory shows that the relative \ch{Dy}--\ch{Fe} spin alignment selects the polarization, while the \ch{Fe} sublattice controls nonrelativistic spin splitting, thus enabling reversible switching.
These results establish spin-induced improper polarization as a route to switchable altermagnetism in nonpolar bulk systems.

\end{abstract}

\date{\today}
\maketitle



\textit{Introduction---}Altermagnets represent a distinct class of collinear magnetism~\cite{2022_PRX_Smejkal, 2022_PRX_Smejkal2, 2024_AdFM_Bai}, combining negligible stray fields and the robustness of antiferromagnets with the momentum-dependent spin-split bands of ferromagnets.
This unique duality makes altermagnets exceptional platforms for next-generation spintronics, enabling phenomena such as spin-polarized transport~\cite{2021_PRL_GHR}, giant magnetoresistance~\cite{2022_PRX_Smejkal3, 2023_PRB_Cui, 2024_PRB_Samanta}, unconventional superconductivity~\cite{2023_PRB_Zhu, 2023_PRB_Brekke, 2024_PRL_Ghorashi}, anomalous Hall effect~\cite{2023_PRB_Nguyen,2023_PRL_Gonzalez,2025_Nat_Zhou,2023_npjCM_Guo}, and other novel spintronic applications~\cite{2023_PRL_Shao, 2024_PRL_Zhang, 2024_PRB_Denisov, 2024_PRL_Jin}.

To realize energy-efficient control of altermagnetism, attention has increasingly turned to multiferroic materials, particularly those that simultaneously exhibit ferroelectricity and altermagnetism~\cite{2023_npjCM_Liu, 2025_PRL_Duan, 2025_PRL_Gu, 2025_PRB_Mavani, 2025_AdvM_Sun, 2025_NL_Wang_magnons, 2025_NL_Wang_EField, 2025_PRL_Zhu}.
Recent high-throughput screening of bulk materials~\cite{2025_PRL_Gu} identified only two ferroelectrically switchable altermagnets among 2001 experimentally reported magnetic structures in the MAGNDATA database~\cite{2016_MAGNDATA}, thus highlighting a severe materials bottleneck.
This scarcity arises from a stringent symmetry constraint in prior design strategies---the need for a polar structure---which greatly restricts the pool of candidate materials.
A natural question is whether a nonpolar lattice can host switchable altermagnetism if the polarization itself is magnetically induced.
Spin-induced improper polarization (SIP)~\cite{2008_JETP_Zvezdin,2010_NJP_Stroppa,2016_PSS_Triguk,2017_NC_Zhao} would offer precisely such a route because it arises from exchange striction driven by magnetic order in an otherwise nonpolar lattice.
In rare-earth orthoferrites, this mechanism operates between the rare-earth and transition-metal magnetic sublattices, giving rise to type-II multiferroicity and nonlinear magnetoelectric responses~\cite{2009_Physics_Khomskii, 2022_PRB_Sasani}.
Here, we ask whether the same dual-sublattice magnetic order can also control the nonrelativistic spin splitting (NRSS).

In this Letter, we propose a symmetry strategy based on SIP to link polarization switching with NRSS reversal in nonpolar dual-sublattice magnets.
Using a simplified \ch{DyFeO3} (DFO) model~\cite{2010_NJP_Stroppa}, we combine symmetry analysis and first-principles calculations to disentangle the roles of the two magnetic sublattices:
the \ch{Fe} sublattice governs the NRSS, whereas the relative \ch{Dy}-\ch{Fe} spin orientation selects the sign of the SIP.
This hierarchy yields a four-domain mapping in the joint order-parameter space of the SIP and the NRSS. 
Reversing the \ch{Fe} moments changes the signs of both the SIP and the NRSS, whereas reversing the \ch{Dy} moments changes only the SIP.
We further examine constrained coplanar magnetic reversal paths connecting collinear spin domains, showing how NRSS and SIP evolve during continuous \ch{Fe} and \ch{Dy} sublattice spin rotations.
These results establish SIP as a route to switchable NRSS without requiring a switchable polar structural degree of freedom, \textit{i.e.}, within a centrosymmetric fixed-ion structural framework.


\textit{Spin-induced improper polarization---}An ideal platform to explore the coupling between SIP and NRSS is offered by perovskite oxides of the form \ch{RMO3} (\ch{R} = rare earth, \ch{M} = transition metal, \textit{e.g.}, \ch{Cr}, \ch{Fe}, \ch{Mn}), which host two magnetic sublattices and represent promising candidates for high-temperature altermagnetism~\cite{2008_JETP_Zvezdin, 2010_NJP_Stroppa, 2016_PSS_Triguk, 2017_NC_Zhao, 2024_AdFM_Bai,2025_Nat_Naka}.
In rare-earth orthoferrites, the electric polarization ($\vec P$) arises from exchange striction between rare-earth and transition-metal magnetic orders~\cite{2010_NJP_Stroppa, 2009_NatM_Tokunaga, 2011_PRL_Xiang, 2012_PRL_Lu, 2022_PRB_Sasani}.
We consider a simplified collinear magnetic model, rather than the full magnetic ground state of DyFeO$_{3}$, in which both the \ch{Fe} and \ch{Dy} sublattices are assigned A-type antiferromagnetic order~\cite{2010_NJP_Stroppa}. 
This minimal model is used to isolate the coupling between SIP and NRSS.

\begin{figure} 
    \centering
    \includegraphics[width=1\linewidth]{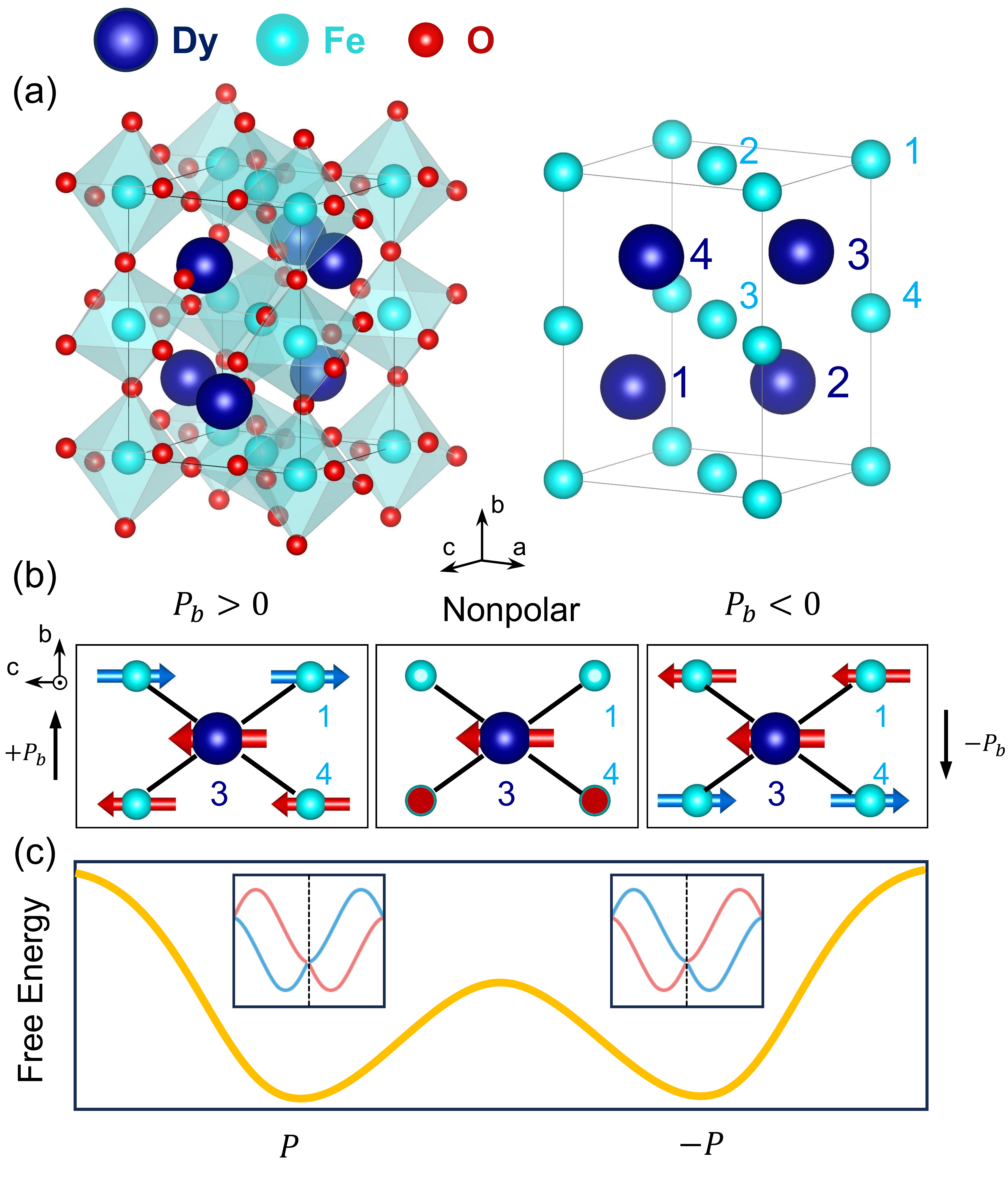} 
    \caption{Polarization and its symmetry coupling to the NRSS in the simplified DFO model.
    (a) Crystal structure of \ch{DyFeO3}, with the rare-earth and transition-metal sites labeled. 
    (b) Parallel and antiparallel (left and right panel) alignments correspond to opposite polarization signs, whereas perpendicular alignment (central panel) gives no net polarization. Red and blue arrows denote opposite spin directions. 
    (c) Schematic illustration of the symmetry-allowed coupling between the SIP and the NRSS.}
    \label{Fig1}
\end{figure}

The asymmetric alignment of \ch{Dy} and \ch{Fe} moments within the $ac$ plane breaks spatial inversion symmetry and allows a polar response along the $b$ axis~\cite{2010_NJP_Stroppa}, as illustrated in Fig.~\ref{Fig1}.
Crucially, this magnetic structure locks the sign of the polarization to the relative orientation of the two magnetic sublattices.
As a result, polarization reversal requires reversing only one of the two N\'eel vectors.

\begin{figure}[t] 
    \centering
    \includegraphics[width=1\linewidth]{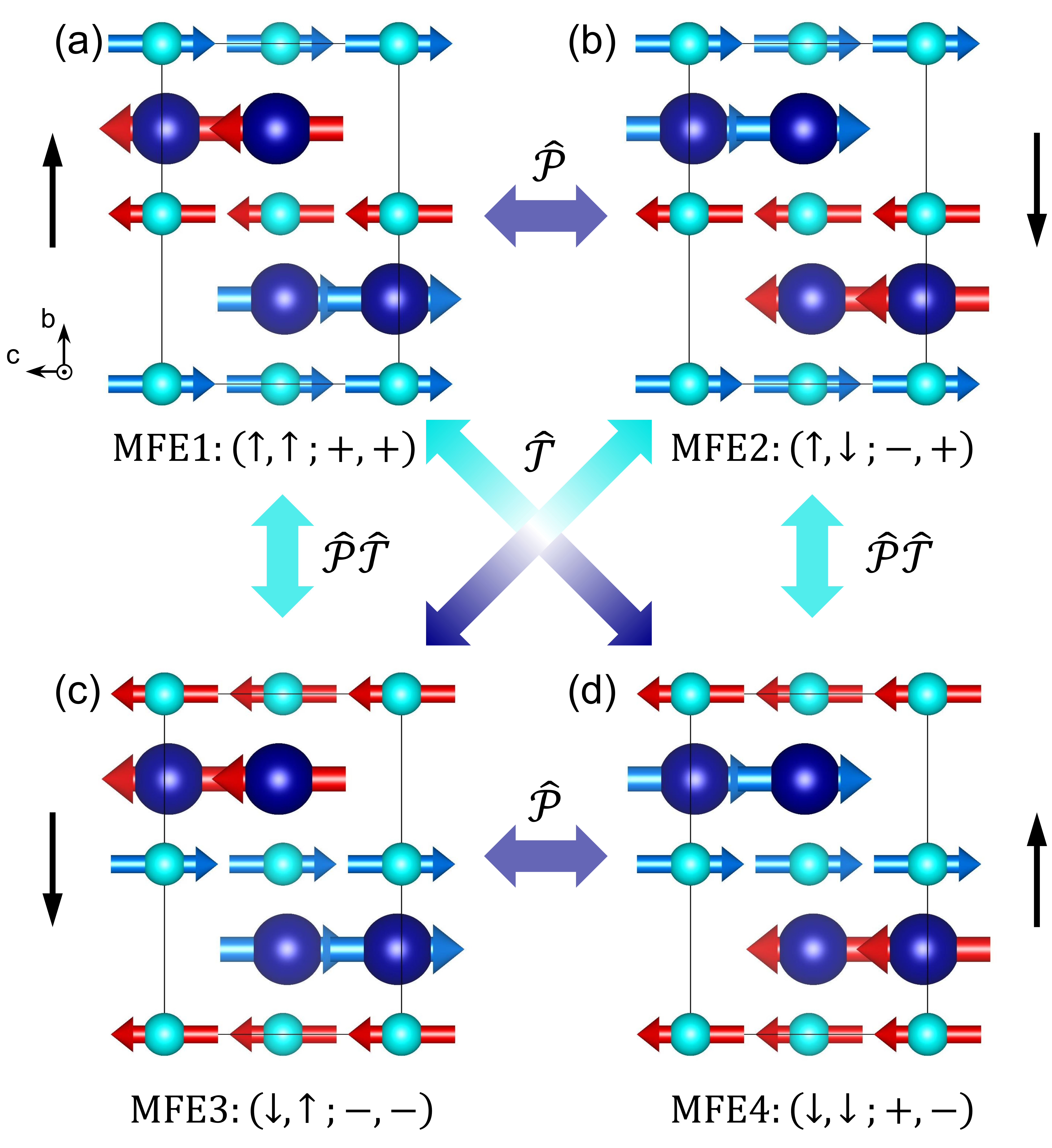} 
    \caption{Symmetry-protected mapping between four multiferroic states. 
    The four states are labeled by $(\vec L_{\mathrm{Fe}},\vec L_{\mathrm{Dy}};P,S)$, in which the first two entries give the orientations of the Fe and Dy N\'eel vectors, and the last two entries give the signs of $P_b$ and $S_{\vec k}$. 
    Red and blue arrows denote opposite spin directions, and black arrows denote the polarization direction.}
    \label{Fig2} 
\end{figure}

Within the described collinear magnetic framework, the system supports four symmetry-related multiferroic states, denoted MFE1--MFE4, with the same polar symmetry setting associated with $Pna2_1$.
First, we introduce a two-component order parameter $(P, S)$.
Here, $P$ denotes the sign of $P_b$, the SIP component along the $b$ axis, and $S$ denotes the sign of the NRSS order parameter, $S \equiv \operatorname{sgn}(S_{\vec k})$, with $S_{\vec k}= \varepsilon_{\vec{k}}^\uparrow-\varepsilon_{\vec{k}}^\downarrow$ in the chosen spin basis.
Second, the multiferroic states can also be described by the orientations of the normalized N\'eel vectors of the two sublattices, denoted by up ($\uparrow$) and down ($\downarrow$) arrows:
\begin{equation}
\vec{L}_{\mathrm{Fe}}\equiv\frac{\vec{m}_{\mathrm{Fe}^3}-\vec{m}_{\mathrm{Fe}^1}}
{2|\vec{m}_{\mathrm{Fe}^1}|},\ \ \ 
\vec{L}_{\mathrm{Dy}}\equiv\frac{\vec{m}_{\mathrm{Dy}^3}-\vec{m}_{\mathrm{Dy}^1}}
{2|\vec{m}_{\mathrm{Dy}^1}|}
\label{subeq:1}
\end{equation}
where $\vec{m}_{\mathrm{Fe}^i}$ and $\vec{m}_{\mathrm{Dy}^i}$ are the magnetic moments on the $i$-th site within the primitive cell labeled in Fig.~\ref{Fig1}(a).
In the DFO model, reversing the \ch{Dy} N\'eel vector while keeping the \ch{Fe} N\'eel vector fixed maps the magnetic configuration onto its spatial-inversion partner ($\hat{\mathcal{P}}$).
Because $\vec P$ is a polar vector, this operation reverses $\vec P$, whereas $S_{\vec{k}}$ remains invariant.
Conversely, reversing the \ch{Fe} N\'eel vector while keeping the \ch{Dy} N\'eel vector fixed maps the configuration onto its combined time-reversal and space-inversion ($\hat{\mathcal{P}}\hat{\mathcal{T}}$) partner.
Under this mapping, both $\vec P$ and $S_{\vec{k}}$ change sign, as summarized in Fig.~\ref{Fig2}.

This behavior can be captured within a Landau framework by the trilinear coupling term:
\begin{equation} 
\mathcal F = \frac{\alpha}{2} P_b^2 + \lambda P_b(\vec L_{\mathrm{Fe}}\cdot \vec L_{\mathrm{Dy}})+\mathcal{O}(P_b^4)
\label{eq:FreeEnergy} 
\end{equation}
Here, $\alpha$ is the polar stiffness, $\lambda$ is the trilinear magnetoelectric coupling coefficient, $P_b$ denotes the $b$-axis component of the SIP, and $\vec L_{\mathrm{Fe}}\cdot\vec L_{\mathrm{Dy}}=\pm1$ for the two collinear parallel or antiparallel configurations.
Minimization with respect to $P_b$ gives $P_b = -(\lambda/\alpha)(\vec L_{\mathrm{Fe}} \cdot \vec L_{\mathrm{Dy}})$, showing that the polarization is an improper secondary order parameter whose sign is determined by the relative orientation of the two magnetic sublattices. Equivalently, the ISODISTORT~\cite{ISODISTORT, Campbell2006} spin-mode decomposition reported in Supplemental Material, Sec. III~\cite{SM_file} separates the MFE1 order into nonpolar Fe- and Dy-centered magnetic modes, \(Q_{\rm Fe}=m\Gamma_1^+\) and \(Q_{\rm Dy}=m\Gamma_4^-\), whose product transforms as the polar mode \(Q_P=\Gamma_4^-\sim P_b\). This provides the symmetry-mode representation of the trilinear coupling in Eq.~(\ref{eq:FreeEnergy}).  
This form is consistent with previous model-Hamiltonian descriptions of rare-earth orthoferrites, where SIP is controlled by the product of rare-earth and \ch{Fe} antiferromagnetic order parameters~\cite{2022_PRB_Sasani}.
We use this coupling as the symmetry basis for linking SIP to NRSS, and show from first-principles calculations that the coupling is governed primarily by the \ch{Fe} sublattice.

\textit{Collinear multiferroic states---}To connect the symmetry analysis to the electronic structure model, we first examine the structural and magnetic symmetries. 
We use the density-functional theory (DFT) framework as implemented in the Vienna ab-initio simulation package (VASP)~\cite{1996_CMS_Kresse, 1999_PRB_Kresse, 1996_PRL_Perdew, 1998_PRB_Dudarev, 1994_RMP_Resta, 1993_PRB_King-Smith, 1992_Fe_Resta}. 
For the symmetry-related collinear magnetic states, we used a constrained-moment approach~\cite{2023_PRB_Chen} to fix the spin directions, enabling a consistent comparison among different magnetic configurations.
All reported values are therefore clamped-ion electronic contributions.

\begin{figure}[t] %
    \centering
    \includegraphics[width=1\linewidth]{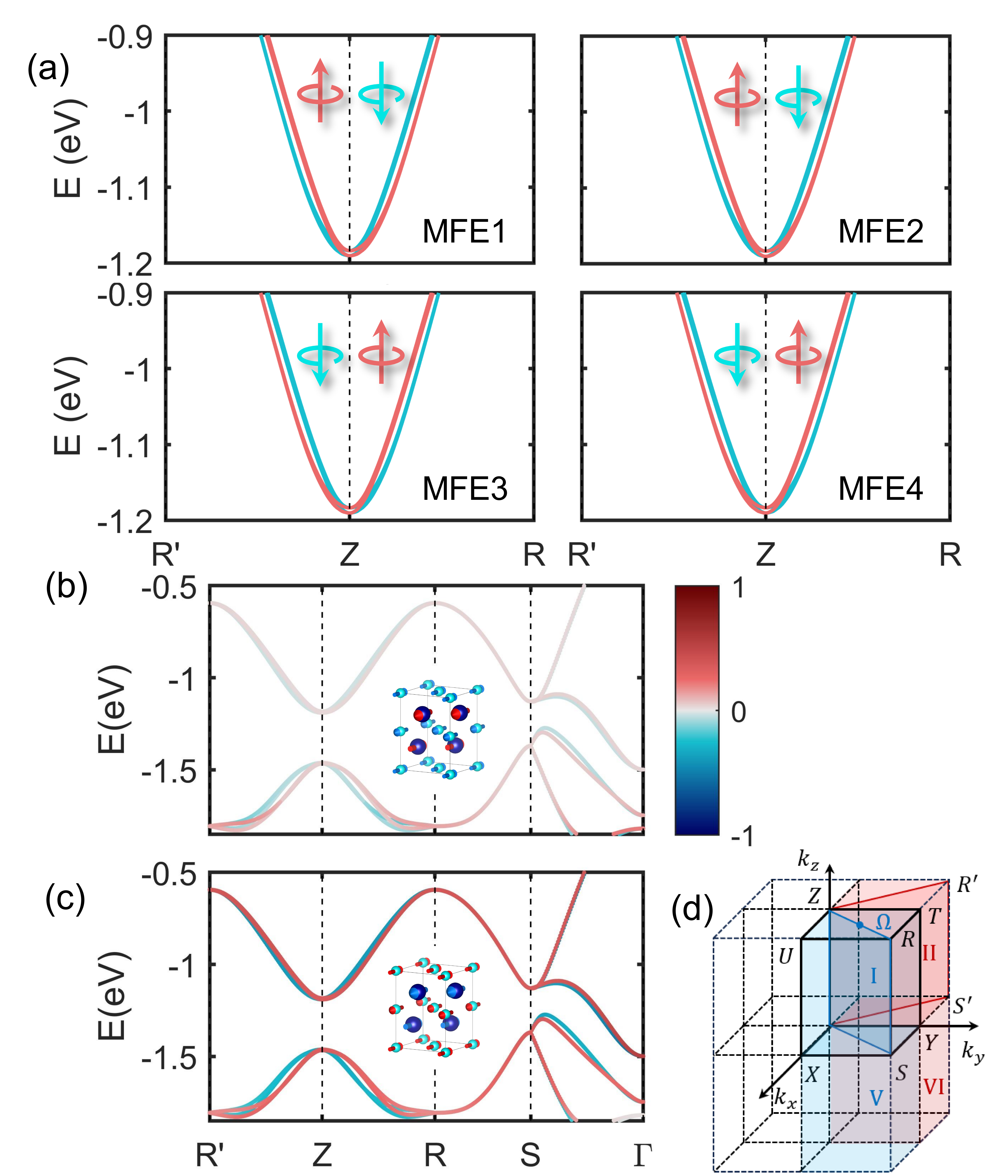} 
    \caption{Electronic structure of the four collinear multiferroic states. 
    (a) Spin-split band structures. 
    Red and blue denote the two spin channels in the chosen spin basis. 
    (b,c) Spin-resolved band structures for MFE1 weighted by Dy (b) and Fe (c) orbital characters, respectively.
    (d) Sign map of the VBM spin-splitting order parameter $S$ in the orthorhombic Brillouin zone. Red and blue denote opposite signs of $S_{\vec k}$, illustrating the altermagnetic sign alternation between symmetry-related momentum sectors.
}
    \label{Fig3} 
\end{figure}

We consider the nonpolar crystallographic space group $Pnma$ (No. 62), which contains inversion symmetry.
When both the \ch{Fe} and \ch{Dy} magnetic sublattices are active, the symmetry reduces to the polar magnetic space group $Pna2_1$.
In the orthorhombic Brillouin zone, we focus on the high-symmetry path $\mathrm{R}'$-$\mathrm{Z}$-$\mathrm{R}$.
At the valence band maximum (VBM), $S$ exhibits a characteristic altermagnetic distribution in which the sign alternates between symmetry-related momentum sectors, as shown in Fig.~\ref{Fig3}(d).
Our first-principles calculations along this path show a maximum spin splitting of about 38 meV for the valence-band edge states.
Crucially, we find that $S_{\mathrm{R'-Z}}^{\mathrm{VBM}}=-S_{\mathrm{R-Z}}^{\mathrm{VBM}}$, a relation dictated by the intrinsic $\left[-1||M_{100}\right]$ spin symmetry, where $[g_s||g_l]$ denotes $g_s$ acting only on spin space and $g_l$ on coordinate space~\cite{2022_PRX_Liu}.
The VBM spin-splitting order parameters of the four multiferroic states satisfy
\begin{equation} 
S_{\mathrm{MFE1}} = S_{\mathrm{MFE2}} = -S_{\mathrm{MFE3}} = -S_{\mathrm{MFE4}}.
\label{eq:S} 
\end{equation}
The calculated clamped-ion SIP components all have magnitude $0.107\ \mathrm{\mu C/cm^2}$, with signs satisfying

\begin{equation} 
P_\mathrm{MFE1} = -P_\mathrm{MFE2} = -P_\mathrm{MFE3} = P_\mathrm{MFE4}.
\label{eq:P} 
\end{equation}
These quantitative results from the DFT calculations are in good agreement with the symmetry analysis in Fig.~\ref{Fig2}.

The \(P\)--\(S\) coupling arises from a trilinear interaction involving the two N\'eel vectors, which induces an improper spin-induced polarization and links it to the \ch{Fe}-sublattice-controlled NRSS identified in this work.
This sublattice hierarchy is further clarified by the limiting symmetry analysis in Supplemental Material, Table~S2~\cite{SM_file}, which isolates the \ch{Fe}-only and Dy-only limits by retaining one magnetic sublattice at a time.
In the Dy-only limit, the relevant spin-space symmetry preserves the spin degeneracy, so the NRSS vanishes; in the Fe-only limit, this protection is lifted, yielding a finite spin splitting whose sign is set by \(\vec L_{\mathrm{Fe}}\).
This interpretation is corroborated by the site-projected, spin-resolved band structures of the MFE1 state [Figs.~\ref{Fig3}(b) and \ref{Fig3}(c)], which show that the VBM states are dominated by \ch{Fe}~\(3d\) character.

Consequently, reversing the \ch{Fe} moments reverses the signs of both \(P\) and \(S\), whereas reversing the \ch{Dy} moments reverses only \(P\), leaving the \ch{Fe}-controlled NRSS unchanged. This sublattice-selective hierarchy provides a symmetry route for coupling polarization reversal to altermagnetic spin-splitting reversal in nonpolar bulk systems.

\begin{figure}[t] 
    \centering
    \includegraphics[width=1\linewidth]{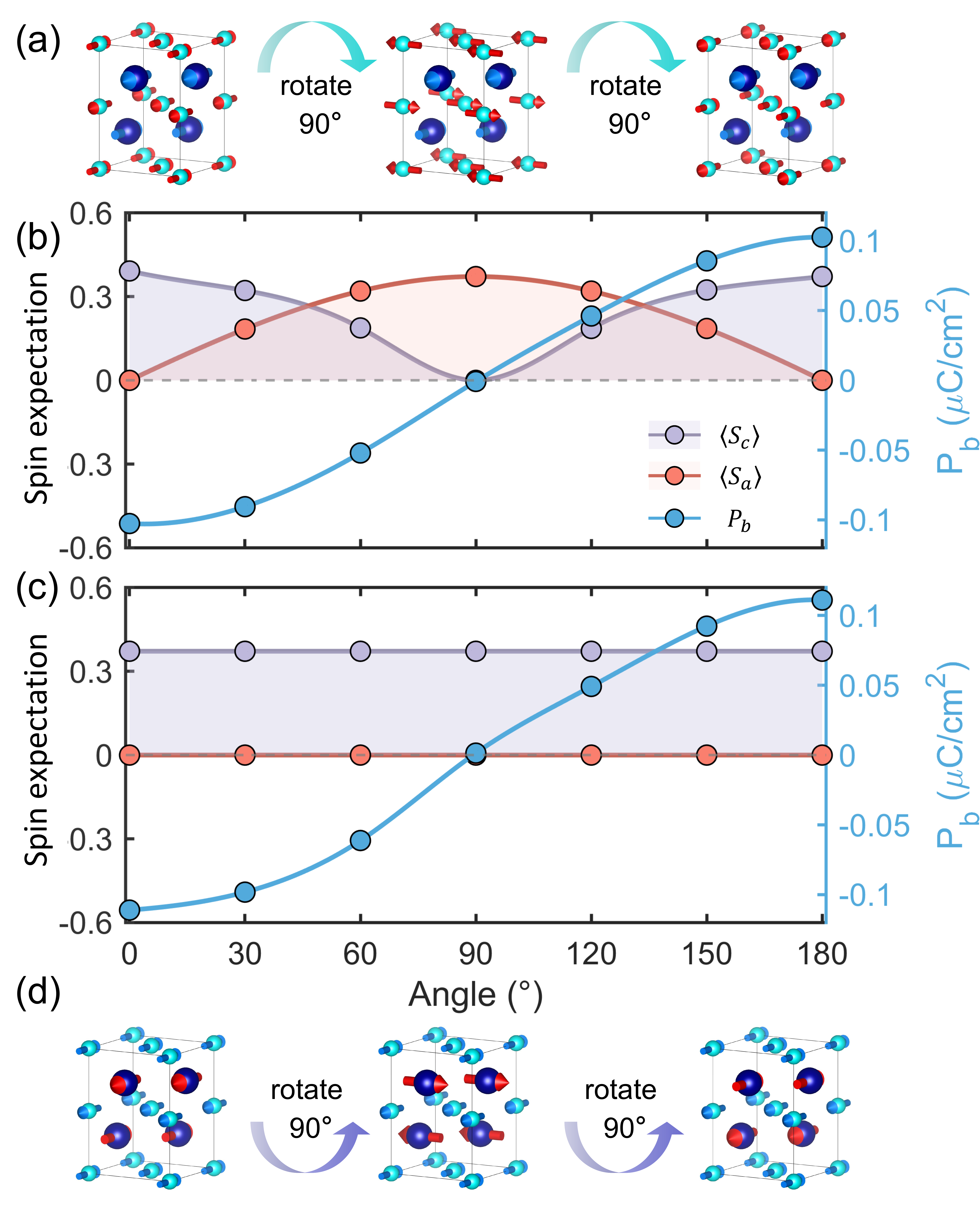} 
    \caption{Constrained magnetic reversal paths.
    (a) Reversal of the \ch{Fe} moments with the \ch{Dy} moments fixed. 
    (b) Evolution of the \ch{Fe}-site spin expectation values $\braket{S_a}$ and $\braket{S_c}$ at $\Omega$, together with the clamped-ion electronic polarization $P_b$, along the \ch{Fe}-reversal path. 
    (c) Corresponding evolution along the \ch{Dy}-reversal path with the \ch{Fe} moments fixed. 
    (d) Schematic Dy-reversal path.}
    \label{Fig4}
\end{figure}

\textit{Coplanar reversal paths}---To illustrate how the coupled order parameters evolve upon spin reversal, we investigate two constrained spin paths in the \(ac\) plane. 
Coplanar magnetic states along the reversal paths were treated using constrained DFT for noncollinear magnetism~\cite{2015_PRB_Ma}. 
Only the magnetic-moment directions were varied along the paths, while the ionic coordinates were kept fixed, consistent with the fixed-ion treatment used for the collinear states. 
Figure~\ref{Fig4} collects the results.
These paths serve to track the order-parameter evolution and are not intended to model the physical switching dynamics, which would require explicit consideration of magnetic anisotropy and spin-orbit coupling.

Path~1 connects MFE1 to MFE3 through a \(180^\circ\) coherent rotation of the \ch{Fe} moments, with the \ch{Dy} moments kept fixed (Fig.~\ref{Fig4}a,b). Path~2 connects MFE1 to MFE2 through the corresponding reversal of the \ch{Dy} moments, with the \ch{Fe} magnetic structure kept fixed (Fig.~\ref{Fig4}c,d).

To monitor the NRSS along the paths, we evaluate the \ch{Fe}-site spin expectation values \(\braket{S_a}\) and \(\braket{S_c}\) at \(\Omega\), the midpoint of the \(\mathrm{Z}\)--\(\mathrm{R}\) segment. As described above, the NRSS at this point directly tracks the \ch{Fe}-sublattice contribution.

Along Path~1, the NRSS remains finite, with its spin character continuously redistributed between \(\langle S_a\rangle\) and \(\langle S_c\rangle\) (Fig.~\ref{Fig4}b). In contrast, Path~2 reverses the clamped-ion polarization \(P_b\) while leaving the electronic bands and NRSS nearly unchanged  (Fig.~\ref{Fig4}c). For an ideal coplanar interpolation, with \(\theta\) the relative angle between \(\mathbf L_{\rm Fe}\) and \(\mathbf L_{\rm Dy}\), the leading exchange-striction invariant gives \(P_b\propto\mathbf L_{\rm Fe}\!\cdot\!\mathbf L_{\rm Dy}\sim\cos\theta\). Thus, \(P_b\) reverses through the orthogonal spin configuration, where the exchange-striction invariant vanishes, while the Fe-derived NRSS remains finite and continuously redistributes its spin projection along the path.

These results establish a consistent microscopic picture of the \(P\)--\(S\) coupling mechanism. The NRSS is governed by the \ch{Fe} sublattice and therefore changes sign when the \ch{Fe} moments are reversed.
In contrast, the sign of the clamped-ion SIP is set by the same sublattice-alignment symmetry relation since the polar response is controlled by the relative orientation of the rare-earth and \ch{Fe} magnetic orders. This domain-reversal analysis should be distinguished from the magnetic-field-driven nonlinear magnetoelectric response of \ch{GdFeO3}~\cite{2022_PRB_Sasani}, in which the field dependence is dominated by the rare-earth sublattice while the \ch{Fe} order remains comparatively rigid.

A key consequence is that a \(180^\circ\) reversal of the \ch{Fe} \(3d\) moments with the \ch{Dy} \(4f\) moments pinned reverses both the polarization and the NRSS. This deterministic sublattice-selective coupling establishes the simplified DFO model as a proof-of-principle platform for coupled polar and altermagnetic responses. A sublattice-resolved force analysis supporting this symmetry interpretation is provided in Supplemental Material, Sec. III~\cite{SM_file}.

\textit{Conclusions---}In summary, using a simplified DFO-derived dual-sublattice model, we have shown that the SIP mechanism known in rare-earth orthoferrites can be coupled by symmetry to nonrelativistic altermagnetic electronic spin splitting. Symmetry analysis and first-principles calculations identify a clear division of roles: the \ch{Fe} sublattice governs the NRSS, while the combined \ch{Dy}--\ch{Fe} order breaks inversion symmetry and selects the improper polarization allowed by symmetry, consistent with the exchange-striction origin of the SIP.

Consequently, reversing the \ch{Fe} N\'eel vector reverses the signs of both SIP and NRSS, whereas reversing the \ch{Dy} N\'eel vector reverses only the SIP. The constrained coplanar paths further show that NRSS remains finite throughout the \ch{Fe}-reversal path, with its spin character redistributed between components. These results establish SIP as an alternative route for linking polarization reversal to altermagnetic spin-splitting reversal in nonpolar systems. As a proof-of-principle symmetry study, this work motivates future material-specific investigations of switchable altermagnetism in type-II multiferroics and related improper polar systems, where polarization is induced by an underlying magnetic order parameter, as in the present case, or by orbital~\cite{OO1,OO2,OO3} or charge order in other materials.

Together with recent demonstrations of N\'eel-vector switching by spin-orbit torques in altermagnetic thin films~\cite{2024_SciAdv_Han} and symmetry-designed all-electric control in bilayers~\cite{2025_PRL_Chen}, our work identifies SIP as a complementary symmetry route to switchable altermagnetism in nonpolar bulk systems, bypassing the restrictive polar-structure criteria that have limited materials discovery.

This perspective promotes dual-sublattice improper polarization from a magnetoelectric effect to a symmetry-based search principle for coupled polar and altermagnetic electronic responses. Looking ahead, the same symmetry principle may be extended to artificial dual-sublattice platforms, such as stacked A-type antiferromagnetic bilayers. In such systems, two layer-resolved N\'eel vectors, \(\mathbf L_1\) and \(\mathbf L_2\), can act as coupled magnetic sublattices; when allowed by the stacking symmetry, a trilinear invariant \(P(\mathbf L_1\!\cdot\!\mathbf L_2)\) can induce a polar mode, with sliding or twisting providing an additional structural control knob.

\textit{Acknowledgments---}We acknowledge financial support from the National Key R\&D Program of China (No. 2022YFA1402901), NSFC (Grants No. 12188101), Shanghai Science and Technology Program (No. 23JC1400900), the Guangdong Major Project of the Basic and Applied Basic Research (Future functional materials under extreme conditions--2021B0301030005), Shanghai Pilot Program for Basic Research---Fudan University 21TQ1400100 (23TQ017), the robotic AI-Scientist platform of Chinese Academy of Sciences, and New Cornerstone Science Foundation. A.S. thanks Paolo Barone, Sang-Wook Cheong, Xiaonong Shen and M. J. Perez-Mato for useful discussions.

\bibliography{reference}

\end{document}